\documentclass[letter]{emulateapj}

\newcommand{\dz}{$\Delta z$ }

\slugcomment{Submitted to ApJ}
\begin{document}

\title{Constraining Extended Reionization Models Through \\ Arcminute-scale CMB Measurements}
\author{Khee-Gan Lee}
\affil{Department of Astrophysical Science, Princeton University, Princeton, New Jersey 08544, USA}
\email{lee@astro.princeton.edu}

\begin{abstract}
The measurement of the optical depth to free electrons, $\tau_r$, in 
the cosmic microwave background (CMB) provides an important constraint
on reionization, but is degenerate to more complex reionization models. 
Small angular-scale CMB measurements of the kinetic Sunyaev-Zeldovich (kSZ) and 
Ostriker-Vishniac (OV) effects have the potential to break this degeneracy. 
We calculate the OV signal for 
various extended reionization histories described by a simple analytic form.
These models are parametrized by \dz, the duration of
the reionization event.  
For reionization histories with identical values of $\tau_r$, 
the OV amplitude at $l = 3000$ ($C_{3000}$) differs by 
$\approx 20 \%$ between the models with 
\dz = 0.1 and \dz = 3.0. 
We demonstrate that the removal of the $z \lesssim 6$ component of the 
OV signal will significantly enhance the ability to differentiate
between  reionization histories, 
with $C_{3000}$ varying by a factor of $\sim 2$ between the models 
with \dz = 0.1
and \dz = 3.0. 
If the low-redshift OV and related kSZ signal can be 
adequately subtracted, this would provide an important observational constraint
on extended reionization models.

\end{abstract}


\section{Introduction}
The first decade of the 21st century has arguably witnessed
 the advent of precision 
cosmology, ushered in by cosmic microwave background (CMB) experiments,
such as the Wilkinson Microwave Anisotropy Probe (WMAP, \citet{komatsu08}), 
and large-scale galaxy surveys, such as Sloan Digital Sky Survey (SDSS, 
\citet{sdss}) and the 
2-degree Field Galaxy Redshift Survey (2dFGRS, \citet{2df}). 
The major cosmological parameters are now known at the percent accuracy level,
and $\Lambda$CDM has emerged as the cosmology which is most easily 
reconciled with observations. 

The current concordance picture of $\Lambda$CDM cosmology holds that after 
the universe became transparent to the CMB
at z $\sim$ 1000, the baryonic content of the universe was neutral 
and coupled gravitationally to the dark matter component. 
Some $10^8$ years after the Big Bang, 
the collapse of massive halos led to the formation of the 
first generation of stars and galaxies. 

These first-generation stars are believed to be extremely massive 
 ($M > 100 M_{\sun}$) and have extremely hard spectra. 
The prodigious amounts of $E \geqslant 13.6\; \mathrm{eV}$ 
photons generated by these
stars propagated out of their host protogalaxies into deep space, 
gradually re-ionizing the inter-galactic medium (IGM), fully ionizing the
universe by $z \sim 6$ (see \citet{barkanaloeb1} and \citet{fo08} 
for reviews on the subject). 
The reionization process is believed to be
highly complex: even though considerable theoretical work (both analytical
and simulations) have been carried out to understand this epoch
(e.g.\ \citet{fzh04,iliev06,trac07}), 
the lack of observational evidence has made it difficult to reach
consensus on fundamental issues such as the temporal extent of
reionization and when it began (e.g.\ \citet{lidz07,shin08}),
 let alone minutae such as 
feedback processes (e.g.\ \citet{ciardi00,mesinger08})
 and the details of ionized bubble percolation (e.g.\ \citet{mcquinn07,lee08})
through the universe. 

At present, there are two major observational constraints on reionization:
(1) The optical depth to free electrons up to the surface of
last scattering has been measured through CMB 
polarization data to be $\tau_r = 0.084 \pm 0.016$ \citep{dunkley08}, 
corresponding to 
a reionization redshift of $ z_r \approx 11$ assuming (unrealistically)
that reionization occured instantaneously.
(2) The detection of the Gunn-Peterson absorption trough from neutral
hydrogen in the Lyman-$\alpha$ absorption spectra of 
high-redshift ($z \sim 6$) quasars indicates a rising
 neutral fraction beginning from this epoch \citep{fan06},
hinting at the end of reionization process, although this is by no means
certain \citep{becker07,furl_mes08}.

Over the next decade and beyond, radio experiments are being 
designed to probe the 21-cm line of neutral hydrogen, 
e.g.\ LOFAR \citep{zaroubi05},
 MWA \citep{morales04},
and SKA \citep{terzian06}. These 21-cm tomography experiments promise full 3D imaging of
the neutral IGM, allowing reionization to be studied in unprecedented 
detail. However, these experiments are still in their early stages.

At time of writing, several high angular resolution CMB experiments
are ongoing, such as the Atacama Cosmology Telescope (ACT, \citet{act}) 
and 
South Pole Telescope (SPT, \citet{spt}). 
These experiments have resolutions approaching
$ 1 '$, with sensitivities of $\sim \mathrm{mK \sqrt{s}}$. The 
temperature-temperature angular power spectra $C_l$ produced by these
experiments will, for the first time, probe small angular scales 
$l \gtrsim 3000$ at which the primordial CMB power spectrum 
 is no longer the dominant contribution to the anisotropy power. 
Instead, the secondary anisotropy at these scales 
will be dominated by foreground
astrophysics such as unresolved radio galaxies, far IR/sub-milimetre 
dusty galaxies and the thermal/kinetic Sunyaev-Zeldovich (tSZ and kSZ, 
respectively) and Ostriker-Vishniac (OV) effects. 

In this paper, we study
the sensitivity of the OV effect towards
extended reionization models with the same value of $\tau_r$.
 While the OV and kSZ effects have been extensively studied in 
the past \citep{ov86,vish87,jk98,hu00,zhang04}, 
these studies were made prior to the WMAP 5th Data Release (WMAP5,
 \citet{dunkley08}) 
constraint on $\tau_r$  (and in the earlier papers,
 before concordance $\Lambda$CDM
cosmology was widely accepted). \citet{mcquinn05} made a study on the OV/kSZ
signal from different reionization models, but this was based on 
preliminary WMAP results, 
and they did not explore the potential of the OV effect to break the 
degeneracy of $\tau_r$ towards different reionization models.

First, we briefly summarize the derivation of the OV effect before
discussing our chosen model for parametrizing reionization histories.
We then present our calculations and discuss them in the context of upcoming
experiments.

\section{The Ostriker-Vishniac Effect}\label{ov_sec}

In this section we summarize the derivation of the CMB anisotropy 
from the OV effect as 
elucidated in \citet{vish87} and \citet{jk98}. 
The fractional temperature change in the direction $\vec{\theta}$
from the scattering of the primordial
CMB by bulk motion of free electrons is
\begin{equation} \label{deltat}
p(\vec{\theta}) \equiv \frac{\Delta T}{T} = 
- \int^{\eta_0}_0 n_e \sigma_T e^{-\tau}
[\hat{\theta} \cdot \mathbf{v}(w\vec{\theta};w)] a(w) dw ,
\end{equation}
where $n_e$ is the electron number density along the line of sight, 
$v(\mathbf{w};w)$ is the bulk velocity at comoving distance $\mathbf{w}$
in units of the Hubble distance,  
at conformal time $\eta = \eta_0 - w$ 
(where conformal time is defined by $d\eta = dt/a$), 
$\sigma_T$ is the Thomson 
scattering cross-section, $a(w)$ is the scale factor of the universe
at $w$ and we set $c = 1$. $\tau$ is the optical depth to 
electron scattering to a given epoch,
 and for a $\Lambda$CDM cosmology it can be
expressed as a function of redshift thusly:
\begin{eqnarray} \label{taueqn}
\tau(z) &=& \frac{\Omega_b \rho_c \,\sigma_T a_0}{m_p} \int^z_0
\frac{x_e(z')(1+z')^2 }{\sqrt{\Omega_m(1+z')^3 + \Omega_{\Lambda}}} dz'
\nonumber \\
&=&\frac{0.046 \Omega_bh x_e}{\Omega_m}
[(\Omega_m(1+z)^3 + \Omega_{\Lambda})^{1/2} - 1]
\end{eqnarray}
where $\Omega_b$,$\Omega_m$ and $\Omega_{\Lambda}$ are the present-day baryon,
matter (dark matter + baryon) and dark energy densities  respectively, 
in units of the critical density $\rho_c$, 
$x_e$ is the ionization fraction and $m_p$ is
the mass of the proton. In the second line, we have assumed that $x_e$ 
is constant over the period of integration.  

The visibility function is defined as
\begin{equation}\label{visfunc}
g(w) = \bar{n}_e(w) \sigma_T a(w) e^{-\tau} = \frac{d\tau}{dw} e^{-\tau}.
\end{equation}
 $n_e$ from Eq.~\ref{deltat} has been replaced by the mean electron
number density $\bar{n}_e = \Omega_b  \rho_c \bar{x}_e(z)(1+z)^3/m_p$. 
$g(w)$ represents the probability distribution 
of first scattering from reionized scattering (contrast the analogous
visibility function in CMB recombination physics, which is the probability
distribution of {\it last} scattering). 
The visibility function is normalized such that 
\begin{equation}
\int^{\infty}_0 g(w) dw = 1 - e^{-\tau_r},
\end{equation}
where $\tau_r$ is the total optical depth to scattering by free electrons
up to the surface of last scattering. 

The fractional temperature change can then be rewritten as
\begin{equation}
p(\vec{\theta}) = - \int^{\eta_0}_0 g(w) \hat{\theta} \cdot
 \mathbf{q}(\mathbf{w}, w) , 
\end{equation}
where $\mathbf{q}(\mathbf{w}, w)=[1 + \delta(\mathbf{w},w)] 
\mathbf{v}(\mathbf{w}, w)$ and $\delta(\mathbf{w},w)$ is the
fractional density perturbation.

\begin{figure*}
\plottwo{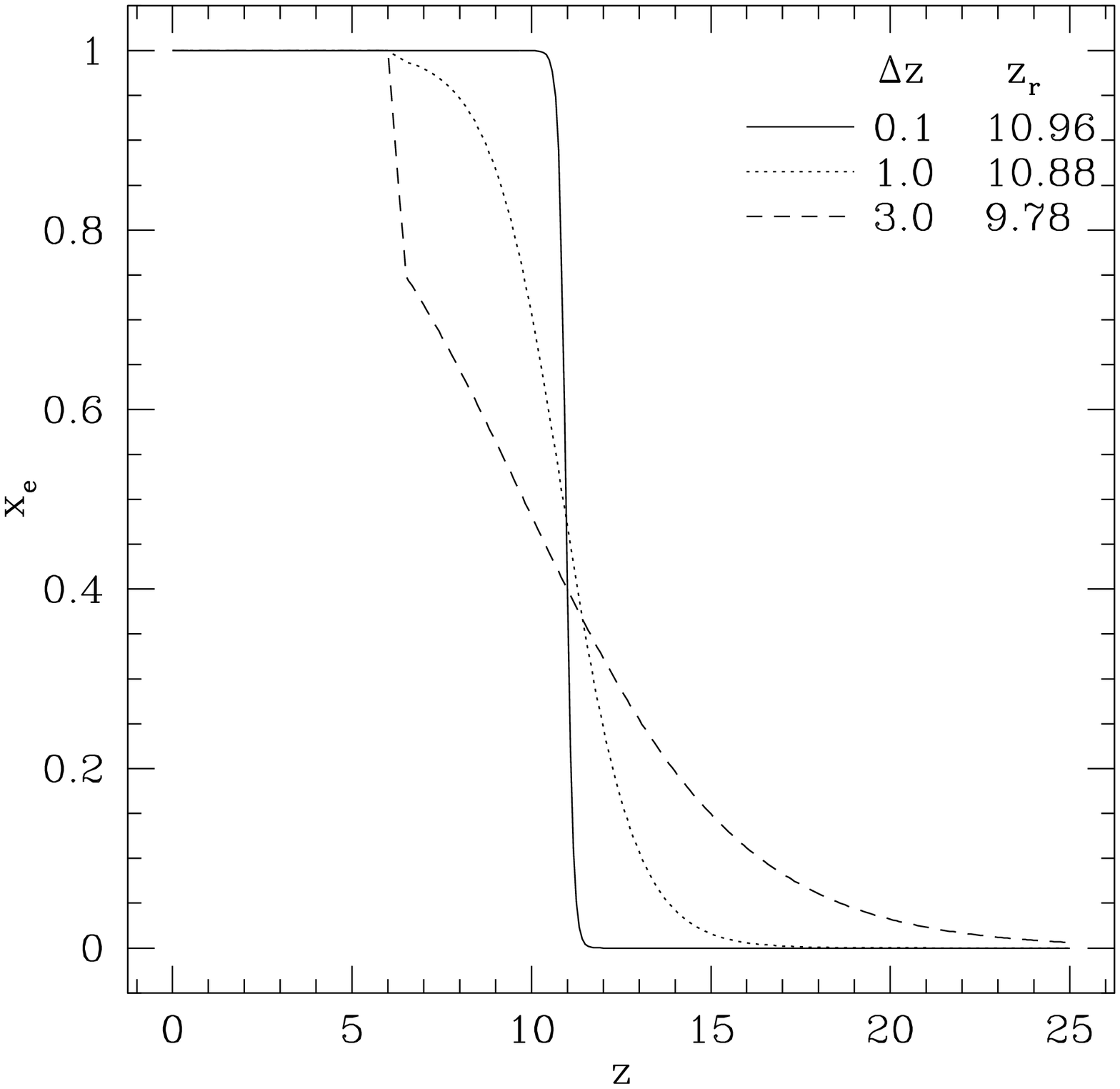}{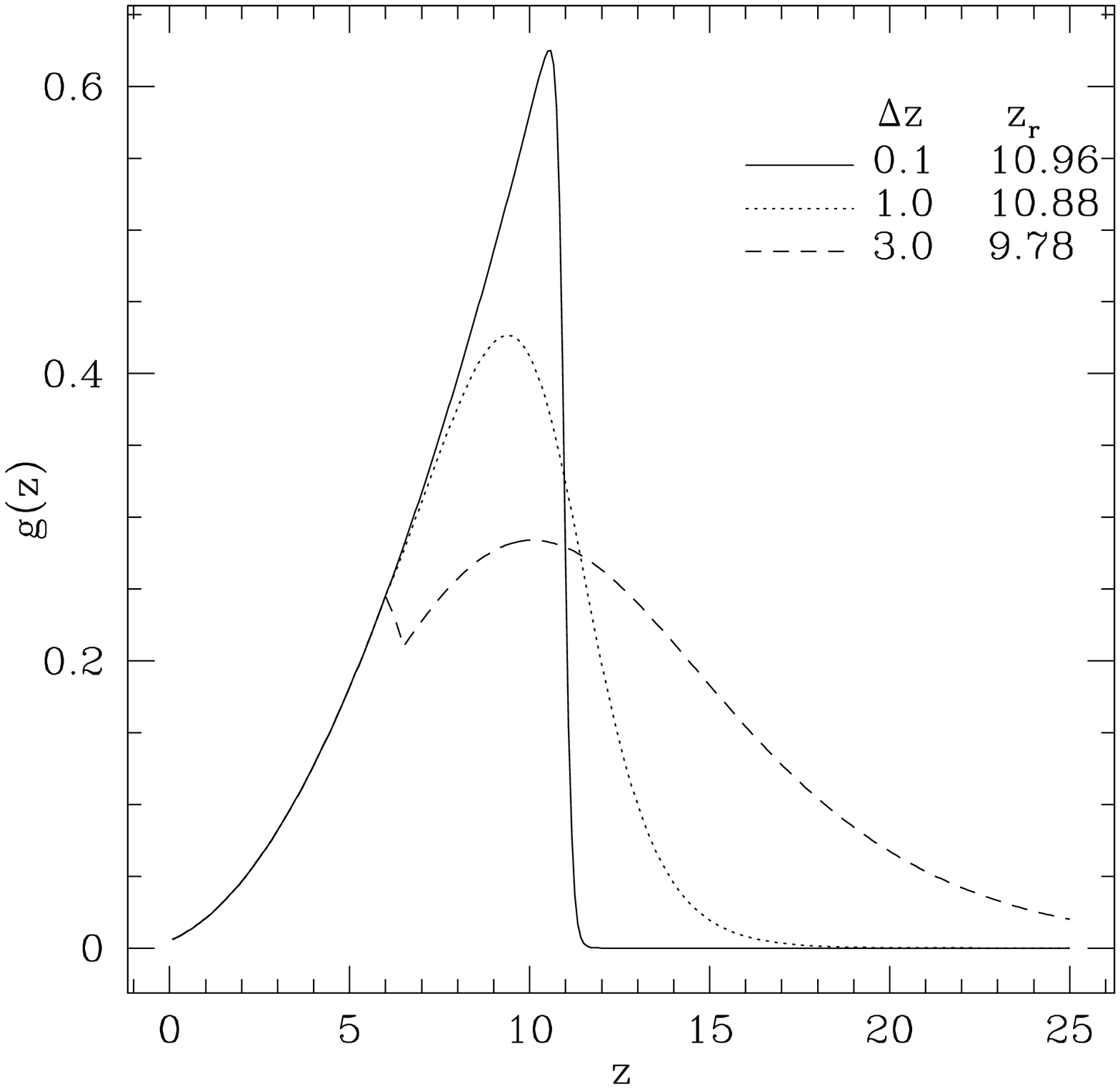}
\caption{Ionization fraction $\bar{x}_e$ (left)
 and visibility function $g$ (right)
as a function of redshift, for 3 reionization histories constrained by
$\tau_r = 0.09$. The discontinuity at $6.0 < z < 6.5$ is caused by our 
linear interpolation scheme which ensures that $x_e(z = 6.0) = 1$
(see discussion in Sec.~\ref{reion_hist}).\label{modelfig} }
\end{figure*}

\begin{figure}
\plotone{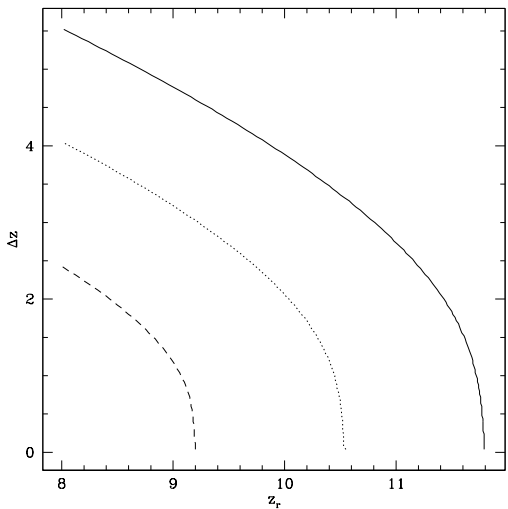}
\caption{Reionization histories parametrized by $z_r$ and $\Delta z $ 
(as described by Eq.~\ref{h1_evol}), solved for $\tau_r = 0.100$
(solid line), $\tau_r = 0.085$ (dotted line) and $\tau_r = 0.070$ 
(dashed line). }
\label{deltazcurves}
\end{figure}

Using the linear theory approximations for $\delta(\mathbf{w},w)$ and 
$\mathbf{v}(\mathbf{w}, w)$, and projecting the temperature perturbation
onto the sky using the Limber approximation, 
we get the Ostriker-Vishniac angular 
power spectrum in terms of multipole moments
\begin{equation}\label{cl}
C_l = \frac{1}{16\pi^2}\int^{\eta_0}_0 \frac{g^2(w)}
{w^2} a(w)^2 \left(\frac{\dot{D}{D}}{D_0^2}\right)^2 
S(l/w) dw,
\end{equation}
where $w = \eta_0 - \eta$, $D$ is the growth factor (the dotted variable
denotes a derivative with respect to time) 
and the function $S(l/w)$ (sometimes known as the 
Vishniac power spectrum) is usually written in terms of
the comoving wavenumber $k$:
\begin{eqnarray}\label{sk}
S(k) &=& k \int^{\infty}_0 dy \int^1_{-1} dx \; P(ky)
P(k\sqrt{1 + y^2 - 2xy}) \nonumber \\
& &\times\frac{(1-x^2)(1-2xy)^2}{(1 + y^2 - 2xy)^2} 
\end{eqnarray}
where $P(k)$ is the standard linear theory matter power spectrum
using the transfer function from \citet{bbks}, normalized to the latest
WMAP5 values.

We now make a brief note on nomenclature regarding 
the related kSZ and OV effects,
which have at times been used interchangeably by authors.
The temperature perturbation equation (Eq.\ \ref{deltat}) was first derived 
by \citet{sz_orig}, and then studied in the context of the 
peculiar motion of individual galaxy clusters in \citet{sz80}. 
\citet{ov86} and 
 \citet{vish87} were the first to study the effect in the context
of large scale structure, specifically with linear perturbation theory . 
We therefore refer to the OV effect as the temperature anisotropy
caused by velocity and density fields in the linear regime, while
the kSZ is its non-linear counterpart (which is the effect that accounts 
for the temperature perturbation from the
 peculiar motion of galaxy clusters).

\section{Reionization histories}\label{reion_hist}

One of the most important outstanding questions on reionization is:
when did it happen?
 
The na\"{i}ve way of answering this is by measuring the total optical depth $\tau_r$
to free electrons from us to the surface of last scattering.
One can then invert Eq.~\ref{taueqn} to find the epoch of reionization $z_r$.
This estimate of $z_r$ assumes an instantaneous reionization event
which is physically unrealistic, while $\tau_r$ by itself is degenerate to more
complicated reionization histories. 

Until recently, measurements of $\tau_r$ have had considerably worse precision, 
and authors studying reionization have tended to regard $\tau_r$ as a tunable
free parameter in their reionization models. 
The WMAP5 measurement\footnotemark of 
$\tau_r = 0.084 \pm 0.016$ 
(corresponding to an instantaneous reionization redshift 
$z_r = 10.8\, \pm\, 1.4$)
has been precise enough to 
change this paradigm, and in this paper we regard $\tau_r$ as a fixed
value, albeit one that still has non-negligible uncertainties.
However,  one can expect subsequent WMAP data
releases as well as the upcoming Planck space mission \citep{planck06} 
to further tighten the constraint on $\tau_r$ to $<10 \% $ precision. 

\begin{figure*}
\plottwo{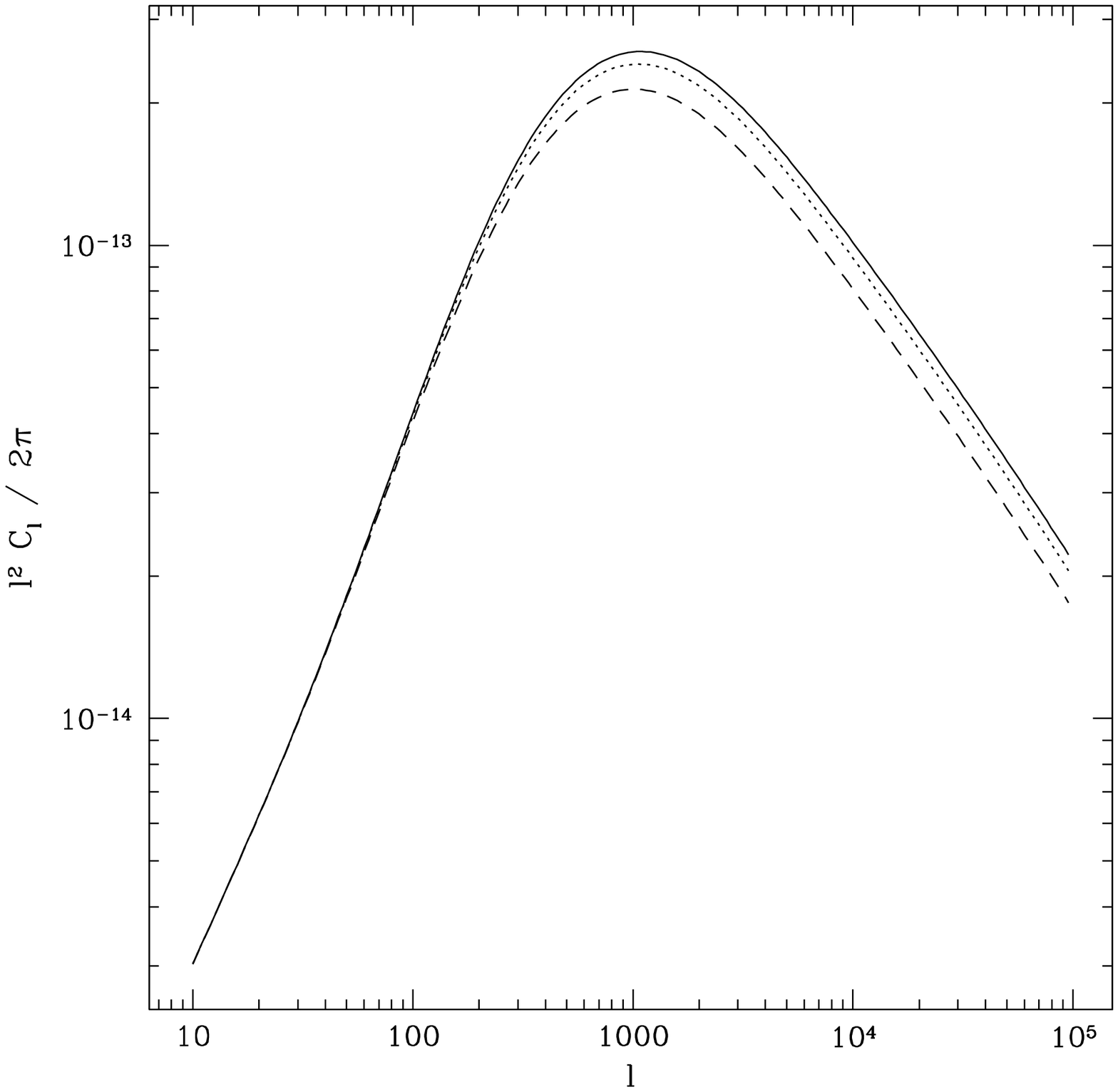}{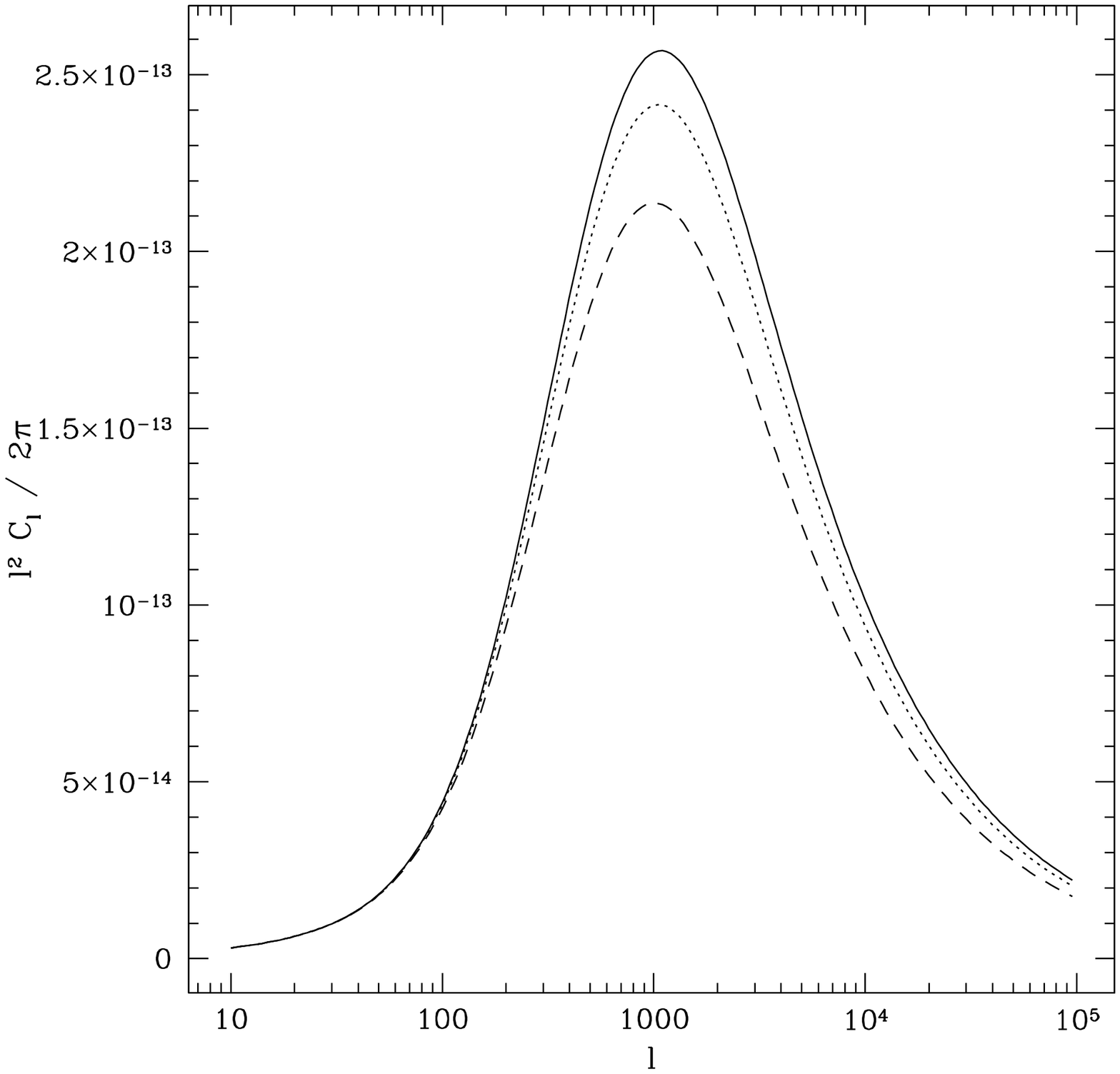}
\label{fullcl}
\caption{Angular power spectra from the Ostriker-Vishniac effect, 
plotted logarithmically (left) and linearly (right). The solid, dotted
and dashed lines correspond respectively to reionization histories
with \dz = 0.1, 1.0 and 3.0. 
}
\end{figure*}

\begin{figure*}
\plottwo{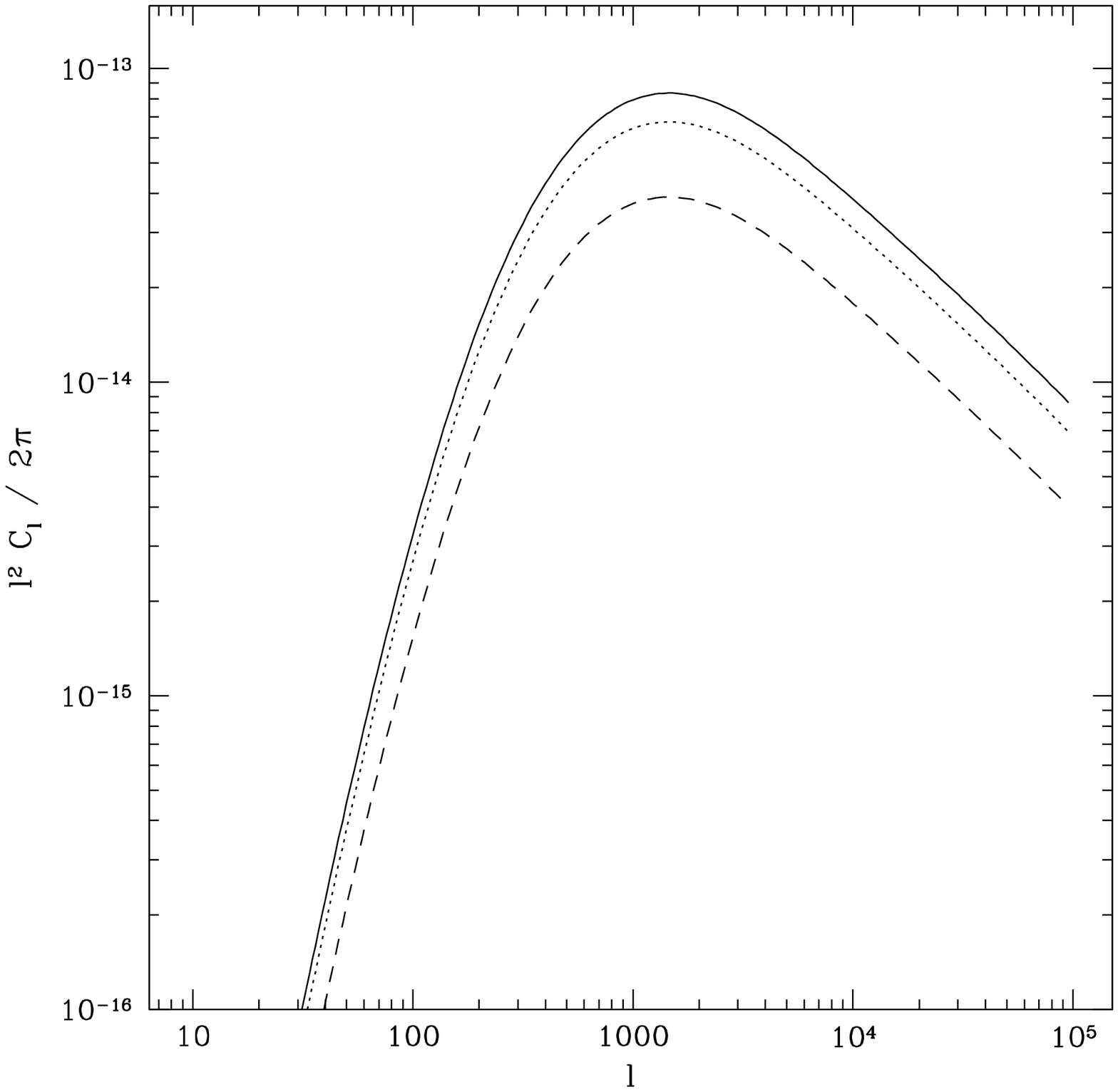}{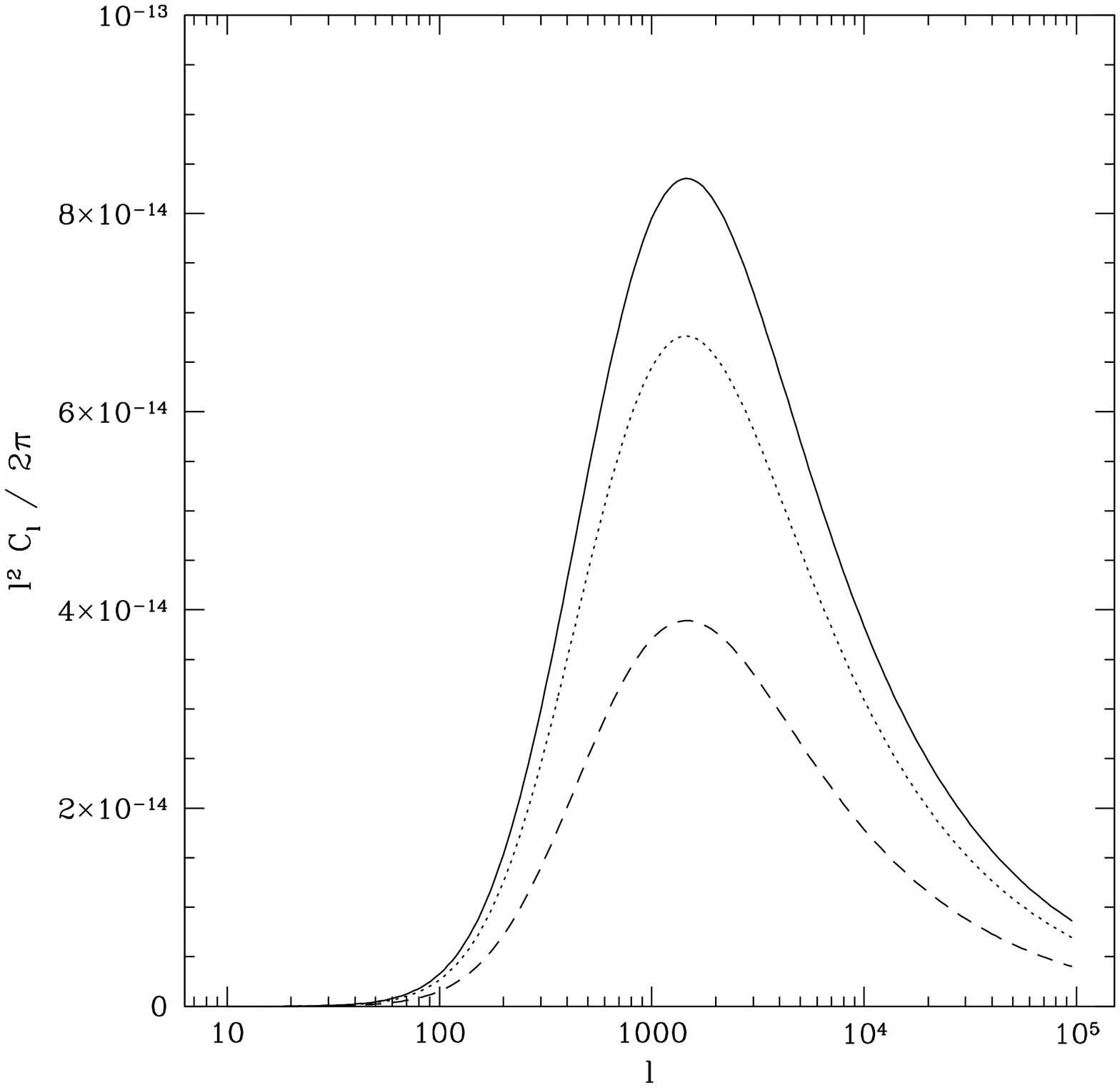}
\label{truncatedcl}
\caption{Angular power spectra from the Ostriker-Vishniac effect 
calculated for $z > 6$, 
plotted logarithmically (left) and linearly (right). The solid, dotted
and dashed lines correspond respectively to reionization histories
with \dz = 0.1, 1.0 and 3.0. Note that the scale on the abscissa differs
from that in Fig.~\ref{fullcl}.
}
\end{figure*}

\footnotetext{http://lambda.gsfc.nasa.gov }

Prior analyses of the kSZ or OV effects have tended to assume instantaneous
reionization, in which the visibility function $g(z)$ has been calculated
assuming that $\bar{x}_e = 1$ at $z \leqslant z_r$ 
and $\bar{x}_e = 0$ prior to that. 

In order to describe an extended reionization event, we choose a simple
parametrization for the
 mean neutral fraction as a function of redshift, often
used to compare results of different reionization models \citep{zfh04,
alvarez06} 
\begin{equation}\label{h1_evol}
\bar{x}_{HI}(z) = \frac{1}{1+ \exp[{-(z-z_r)/\Delta z}]},
\end{equation}
where $z_r$ in this case is defined as the redshift at which 
$\bar{x}_{HI} = 0.5$ and $\Delta z$ parametrizes the duration of the reionization
event. The mean ionized fraction is then trivially $\bar{x}_e = 1 - \bar{x}_{HI}$. 

Up to this point, we have assumed that the universe is composed
solely of hydrogen, and ignored helium reionization. 
It is often assumed that the \ion{He}{1} in the universe 
becomes singly ionized at roughly the same time
as \ion{H}{1} as their (first) ionization energies 
are of the same magnitude (24.6 eV for
\ion{He}{1} compared with 13.6 eV for \ion{H}{1}).
To make this assumption, we modify the expression for the number density
of electrons used in Eqs.~\ref{deltat}--\ref{visfunc}
\begin{equation}\label{he_numden}
\bar{n}_e = \left[\frac{1 - Y}{m_p} + \frac{Y}{4 m_p} \right]
\Omega_b  \rho_c \bar{x}_e(z)(1+z)^3
\end{equation}
where $Y$ is the helium mass fraction of the universe 
(in our calculations we use the value $Y = 0.24$).
 The two terms in the square parentheses correspond to the free electrons
contributed by the ionization of  hydrogen and the first ionization
of helium, respectively. 

\ion{He}{2} reionization, on the other hand,  is a low-redshift ($z \sim 3$) event 
which is much more amenable to observations \citep{shull04,fo08} 
and thus beyond the scope of 
this paper.

While the functional form of Eq.~\ref{h1_evol} is believed to describe well 
the evolution of the neutral fraction
during most of the reionization process, 
it becomes less accurate near the 
end of reionization as $\bar{x}_{HI} \rightarrow 0$.
For reionization histories 
consistent with $\tau_r \approx 0.09$ and $z_r \approx 10$ 
the neutral fraction doesn't 
drop quickly enough to  $\bar{x}_{HI} \sim 10^{-3}$ by $z \lesssim 6.5$, as
observed in high-redshift quasar observations. 
We thus modify our reionization
model as follows:
the neutral fraction is 
 parametrized by Eq.\ \ref{h1_evol} in the regime $z > 6.5$, 
and then interpolated linearly to $\bar{x}_{HI} = 0$ at $z = 6.0$.
The left-hand plot of Fig.~\ref{modelfig} illustrates our chosen form
of $\bar{x}_e$. 

For a fixed $\tau_r$, this gives a family of reionization
models parametrized by $z_r$ or $\Delta z$ . 
In Fig.~\ref{deltazcurves} we compute
$\Delta z $ and $z_r$ for 
reionization histories constrained by 3 different values of $\tau_r$
consistent with WMAP5 (hereafter we use \dz to label 
different reionization histories at fixed $\tau_r$).
 We see that any given value of $\tau_r$ 
is degenerate with a wide range of reionization histories ranging 
from instantaneous ($\Delta z \rightarrow 0$) to highly extended
($\Delta z \sim 4-5$). Note that smaller values of $\tau_r$ tend to limit the possible
range of \dz 
due to our constraint that reionization completes by $z \sim 6$, as 
discussed above.

 Fig.~\ref{modelfig} shows our chosen form of the 
ionization fraction $\bar{x}_e$ and the corresponding 
visibility function $g$ as a function of
redshift, calculated for several reionization histories which
share the same value of $\tau_r$. 

\section{Angular power spectrum calculations}

Eqs.~\ref{cl} and \ref{sk} are integrated numerically to calculate
the OV angular power spectrum. 
In our calculations we use the WMAP5 recommended values \citep{dunkley08}
for the 
cosmological parameters, $h = 0.71$ 
(where $H_0 = 100 h \: \mathrm{km\, s^{-1}\, Mpc^{-1}}$), 
$\Omega_m = 0.28$, $\Omega_b h^2 = 0.0227$, 
$n= 0.96$, 
$\Delta_R^2 (k = 0.002 \mathrm{\:Mpc^{-1}}) = 2.41 \times 10^{-9}$ . 

However, dark energy drops to less than $10\%$ of the total 
mass-energy of the universe above $z \simeq 1.3$,
while the main epoch of interest at which reionization affects the OV signal 
is $ z \gtrsim 6$.  
We thus make the simplifying
assumption of a flat Einstein-de Sitter universe ($\Omega_{\Lambda} = 0$,
$\Omega_m = 1$) for computing the OV angular power spectrum.
 This allows us to make the substitutions 
$a = \eta^2/\eta^2_0$ and $D\dot{D}/D^2_0 = 2 \eta/\eta_0$ in Eq.~\ref{cl}.
The integrated optical depth $\tau(z)$, however, is highly sensitive to 
cosmology and we assume $\Lambda$CDM cosmology when calculating this
for the visibility function (c.f.\ Eq.~\ref{visfunc}).   

While comparing the results of our calculations 
with those in \citet{jk98}, we found discrepancies
that were eventually traced to an error in their 
code for calculating $S(k)$ (c.f.\ Eq.~\ref{sk}). 
This error caused an overall underestimate (by
$\sim 10-20 \%$) in the OV power 
of the results presented in their paper.  
The shape of their $l^2 C_l$ power spectra should
 also be slightly more peaked, 
and the peaks have tended to move to lower $l$ after the
mistake was corrected. 
Hence, 
the analytic approximation for $S(k)$ 
fitted to their results (Eq.~41 in their paper)
is incorrect. 
Nevertheless, the equations derived in their paper are mostly correct, 
except for their Eq.~38, which should be identical to the second line
of our Eq.~\ref{taueqn}.

The resultant power spectra for several reionization histories 
consistent with $\tau_r = 0.09$ are shown in Fig.~\ref{fullcl}.
The power spectra peak at $l \simeq 1000$ while the variance of the 
temperature distribution
\begin{equation}
\langle \left( \frac{\Delta T}{T} \right)^2 \rangle
= \frac{1}{2\pi} \int l\, C_l\: dl 
\end{equation}
corresponds to a root-mean-squared (r.m.s.) 
value $\Delta T \simeq 2.4-2.7 \: \mathrm{\mu K}$.
The amplitude of the spectra tends to decline with increasing
$\Delta z$, as this has the effect of spreading $g(z)$ to higher redshifts 
(c.f.~Fig.~\ref{modelfig}) while the $(aD\dot{D}/wD_0)^2$ parts
of the integrand are decreasing functions with redshift. 
Hence, a more extended reionization event 
 samples the low-redshift part of the integral
less than a rapid event, which decreases the power in the signal.

 At $l = 3000$, where the primary anisotropy drops below the level 
of the secondary anisotropies, $C_{3000}$ differs by about $20 \%$ between
the cases of near-instantaneous ($\Delta z = 0.1$) and highly extended
($\Delta z = 3.0$) reionization. 

However, for the majority of the universe's lifetime we know 
the ionization state: $\bar{x}_e = 1$ from $z \lesssim 6$ onwards irrespective
of the reionization model, so if we remove the 
low-redshift part of the integral we get rid of the part of the signal
that is independent of the reionization model. 

\begin{figure}
\plotone{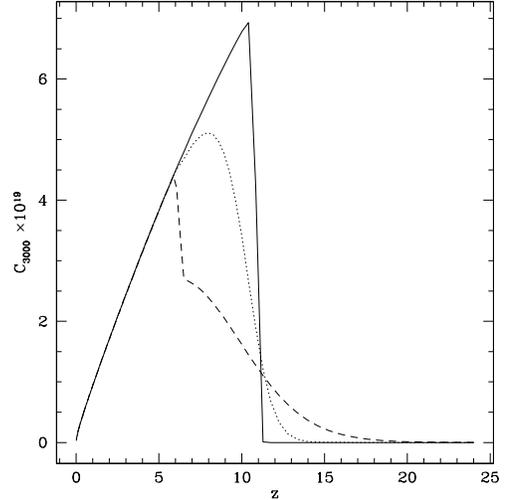}
\label{plotint}
\caption{OV anisotropy at $l = 3000$, plotted as a function of the contributing
redshift for \dz = 0.1 (solid), \dz = 1.0 (dotted) and \dz = 3.0 (dashed). 
}
\end{figure}

In Fig.~\ref{plotint} we plot the contribution of different redshifts
in the calculation of $C_{3000}$ (i.e.\ the integrand of Eq.~\ref{cl}
plotted against redshift for $l=3000$). We see that a significant
portion of the power is contributed from $z<6$.
Removing this contribution would increase the 
contrast between the $C_l$'s computed for
different reionization histories.

This is duly shown in Fig.~\ref{truncatedcl}, where we have removed
the $z < 6$ portion from the integral in Eq.~\ref{cl}. 
The remaining power in the anisotropies peak at $l \simeq 1500$, 
but now the r.m.s.\ temperature perturbation varies from 
$\Delta T = 1.5\: \mathrm{\mu K}$ for $\Delta z = 0.1$ 
to $\Delta T = 1.0 \: \mathrm{\mu K}$ for $\Delta z = 3.0$.
$C_{3000}$ now varies by a factor of $\sim 2$ between the near-instantaneous
reionization model and the most extended one. 

\section{Discussion / Conclusion}\label{conclusion}
From our calculations, we have shown that by altering the duration of the
reionization event that goes into the visibility function of the 
Ostriker-Vishniac power spectrum, significant changes are made to the
resulting $C_l$'s. However, 
the OV effect is certainly not the only secondary anisotropy 
to exist at $l > 10^3$. 

Assuming that instrumental noise issues have been dealt with, 
the main challenges facing the detection of the kSZ/OV signal at $l > 2500$
are the other secondary anisotropies. The galactic foregrounds and
far-IR/sub-mm extragalactic sources can be removed by cross-correlating 
with multi-wavelength data, while the thermal SZ signal can be removed
due to its spectral dependency near the CMB peak temperature. 

In this paper we have only considered
the linear OV effect, and not the full combination of  OV
(which dominates at high-$z$)  and
non-linear kSZ (which dominates at low-$z$).
 However, the calculation of kSZ is carried out through an
integral similar in form to Eq.~\ref{cl}, with the main differences being
the calculation of the integral $S(k)$ (Eq.~\ref{sk}).
In practice, these involve using some form of non-linear 
power spectrum instead of the linear $P(k)$ we used in Eq.~\ref{sk}
\citep{hu00,ma_fry02}. 
These calculations are still weighted by the electron visibility function
$g(z)$ in the same manner as the OV, but the fact that the non-linear terms 
are relatively small compared to the linear terms at $z > 6$ mean
that the kSZ is less sensitive to the reionization history than the OV;
see \citet{carlos_ho09} for a recent paper on this. 
One additional source of 
OV/kSZ power is from 
patchy reionization, i.e.\ the
contribution from large ionized \ion{H}{2} bubbles which form during reionization
\citep{mcquinn05}. 
However, this contribution should similarly be weighted by the visibility
function and would not qualitatively affect the dependency of the kSZ/OV
signal on the reionization history as discussed in this paper. 

In the most optimal case, the noise sensitivity of ACT would allow for
 $C_l$ measurements at a precision of several percent at $l$ of several 
thousand \citep{act}. 
Once the systematic errors from the subtraction of foregrounds,  
extragalactic point-sources and tSZ are taken into account, it would
be challenging to measure the full kSZ/OV signal to sufficient accuracy
to distinguish between different reionization histories, for which the values
of $C_{3000}$ vary by about $20 \%$ between the cases of 
instantaneous reionization and highly extended (\dz = 3) reionization. 

The key to distinguishing between reionization histories lies in removing
the low-redshift ($z < 6$) component from the kSZ/OV signal
(specifically the kSZ contribution from non-linear structures, 
which dominate at low redshift). 
This would greatly increase the contrast between the different reionization
histories, to a factor of $\sim 2$ between the cases of
\dz = 0.1 and \dz = 3.0 as we have seen above.

\citet{ho09} have developed a method of measuring the
kSZ signal at low redshift. 
They build a template of the momentum field by reconstructing
the momentum field from galaxy redshift surveys, 
and then cross-correlating this with arcminute-scale CMB
data. For a cross-correlation between a 4000 sq.\ deg.\
ACT experiment with galaxy survey data from the 3rd phase of the SDSS (SDSS3), 
\citet{ho09} estimate a kSZ detection with a signal-to-noise (S/N) of $ \sim 15$.
However, this method can only measure the kSZ contribution 
from relatively low redshifts ($z \lesssim 1$) that can 
be adequately covered by the galaxy redshift surveys of 
the foreseeable future. 

One possibility would be to construct a kSZ template 
from the Ly$\alpha$ forest in the spectra of high-redshift quasars, 
which coincidentally covers  the
$z \lesssim 6$ redshift region for which we want to subtract the kSZ.
This would be complicated by the difficulty of extracting the 
underlying density field from Ly$\alpha$ data, but even a relatively
low S/N 
 estimate of the low-redshift kSZ should allow us
differentiate between instantaneous and highly extended reionization. 
Considering the current paucity of observational data on 
reionization, even a rough estimate of \dz would provide 
valuable constraints on reionization.

\acknowledgments 
The author thanks David Spergel and Andrei Mesinger 
for useful discussions and advice, 
as well as Marc Kamionkowski and Andrew Jaffe 
for kindly providing access to their
code from \citet{jk98}.

\bibliographystyle{apj}
\bibliography{ms}
\end{document}